\documentclass[conference,a4paper]{IEEEtran}
\ifCLASSINFOpdf
  \usepackage[pdftex]{graphicx}
\else
\fi
\usepackage{pifont}
\usepackage[font=small,labelfont=bf]{caption}

\DeclareRobustCommand*{\IEEEauthorrefmark}[1]{\raisebox{0pt}[0pt][0pt]{\textsuperscript{\footnotesize #1}}}

\begin{document}
%
\title{An Intelligent Mobile Application to Monitor and Correct Sitting Posture Using Raspberry Pi and MediaPipe Pose Detection}

\author{\IEEEauthorblockN{
Yung-Chen (Hailey) Hsieh\IEEEauthorrefmark{1}
}                                     
\IEEEauthorblockA{\IEEEauthorrefmark{1}
Pacific American School, Hsinchu City, Taiwan
}

\IEEEauthorblockN{
Yu Sun\IEEEauthorrefmark{2}
}                            

\IEEEauthorblockA{
\IEEEauthorrefmark{2} California State Polytechnic University, Pomona
}
}



\maketitle

\begin{abstract}
Poor posture has become an increasingly prevalent concern due to students and workers spending extended amounts of time sitting at a desk. To address this issue, we developed \textit{PoseTrack}, a mobile application that uses a Raspberry Pi Camera and Mediapipe Pose landmarks to monitor the user’s posture and provide real time feedback. The system detects poor posture, including forward lean, slouching, hunched shoulders, crossed legs, etc. Some challenges we faced were obtaining posture data, transferring data from the Raspberry Pi to the App, and safely storing user data. We used a Flask server to pass data from the Raspberry Pi to the mobile application, Firebase to store user data, and the Flutter framework to create the app. To test the analysis system’s viability, we designed an experiment that tested the system’s accuracy across several different perspectives and postures. The results indicate that the system is able to effectively detect poor posture whenever the user’s joints are not blocked by the table or their limbs. The results demonstrate the potential for the system to be further improved and used on a larger scale for poor posture monitoring. 
\end{abstract}

{\smallskip \keywords Posture Detection, Raspberry Pi, Mediapipe, Flutter.}

%
\IEEEpeerreviewmaketitle
\thispagestyle{plain}

\vspace{7pt}
\section{Introduction}
After the COVID-19 pandemic, companies and schools are increasingly shifting to remote work. Students and workers now spend more time at home than in physical classrooms and offices, leading to less supervision and exacerbating worldwide issues of poor posture \cite{Espinola}. A study conducted by Susilowati et al. (2022) revealed that after a university work-from-home program, up to 70\% of individuals surveyed reported pains in the neck, lower back, and shoulders \cite{Susilowati}. These symptoms were reported alongside poor postures such as sitting at a table for extended periods, prone positions while using a phone, and latent positions when using a laptop \cite{Susilowati}, highlighting the significant effects of poor posture on muscle pain \cite{Yoo}.  
Poor posture doesn’t merely cause muscle pain but also negatively affects human physical function. Lei Lu et al. (2020) found that only 15 minutes of sitting in poor posture results in a decreased ability to do push ups \cite{Lu}, indicating that sitting in poor posture for a short time can significantly harm human back muscles. Moreover, long term poor posture lowers the tolerance of muscle tissues, causing tissue damage and resulting in musculoskeletal disorders (MSDs) \cite{Anghel}. According to Dong et al. (2022), lower back MSDs increase the amount of inflammatory cytokines in the body and lower the individual’s pain threshold, thereby increasing pain sensitivity compared to healthy individuals \cite{Dong}. Students and office workers are especially prone to such disorders due to performing long hours of work at a desk. If left unaddressed, poor posture can result in chronic Musculoskeletal disorders and lower quality of life, underscoring the importance of correcting poor posture.\\

In this paper, we propose an app called \textit{PoseTrack} to assist students and workers in correcting their postures. We first obtain camera inputs from a Raspberry Pi 5, then analyze the video frames using Mediapipe pose landmarks to determine whether the user is sitting in the correct posture. After the camera input is analyzed, the information is sent and displayed on the app interface. Whenever the user is sitting in poor posture, a warning sign would pop up on the app interface and the speaker would play an alert sound, reminding the user to correct their posture.\\

Our solution is effective because it provides real-time feedback and does not include a wearable device, which makes it more accessible to the public and does not restrict the user’s movement. Instead of focusing only on one part of poor posture, the system provides a comprehensive check that includes crossed legs, slouching, head tilt, and more. This makes the system a more thorough assessment of posture. In addition to the warning pop-up on the mobile app, our system provides haptic feedback from the speaker that can warn the user to change posture even when they’re concentrating on their work and not paying attention to their phone. In addition, our system displays a graph for users to view past trends, which allows users to see how their posture has improved and motivates users to be mindful of their posture. Overall, \textit{PoseTrack} employs a simple camera and mobile application, enabling a user-friendly and non-intrusive monitor of poor posture.\\

We conducted 2 experiments to test the performance of the posture analysis system. The first experiment tested the system’s ability to accurately identify four postures: good posture, forward lean, crossed legs, and leg-on-chair. The results showed that the system accurately identified good posture and forward lean, but had lower accuracy in detecting leg positions due to certain joints being occluded by the table or user arms. The second experiment tested the performance of the analysis system in environments with different lighting. We tested good posture and forward leaning in 5 different lighting conditions while all other environmental factors were kept constant. The results revealed a high accuracy for all the videos except a consistent inaccuracy for the left diagonal video due to occlusion of the user’s legs from the table. This suggests that the system has the potential to be used in various lighting conditions as long as the position of the user joints is in clear camera view.\\

Tlili et al, Gupta et al, and Jang et al. \cite{Tlili}\cite{Gupta}\cite{Jang}  all used wearable devices such as accelerometers or belts with sensors to detect bad posture, which would be difficult to use outside a laboratory setting. The out method eliminates the need for a wearable device, which can make the posture detection application more widely accessible to the general public. The three prior methodologies applied various machine learning techniques—including support vector machines, k-means clustering, and neural networks—to classify posture as good or poor, each achieving approximately 90\% accuracy. In contrast, our approach does not rely on custom-trained machine learning models; instead, we utilize MediaPipe’s pose detection framework to analyze the user’s posture in real time. A key limitation of the previous methods is their narrow focus—typically addressing only a single aspect of poor posture, such as slouching or head and shoulder alignment. Our solution broadens the scope by incorporating multiple posture indicators, including crossed legs, backward leaning, forward head position, and feet placement, resulting in a more comprehensive and holistic posture evaluation.

\vspace{7pt}
\section{Challenges}
\subsection{Obtaining Posture data}
The main function of our program is to accurately detect whether the user is sitting in the correct posture. A challenge is to obtain posture data so that we can analyze the user’s joint positions. The program needs to correctly identify the user’s joints and body positions despite challenges with the user’s clothing and the video background, such as the environment lighting and camera angle. To address this challenge, I could use the mediapipe library’s pose detection model, which helps identify key body locations in images. Mediapipe pose detection is trained on diverse datasets and can identify body joints despite environmental variations.
\subsection{Transfer information from the device to the Application}
Another challenge is to reliably pass the data from the Raspberry Pi device to the app so that the result of the analysis can be displayed. We need to ensure that the app is getting data from the correct device and that the app and the device can interpret the data format. To address this, we can use a Flask server that the app can request information from using HTTP. The app would send a request to the server to validate the url response and retrieve the result of the analysis. The analysis result could be standardized and sent as JSON over HTTP so that the Python analysis and Dart application can both interpret the result.
\subsection{Storing User Data}
In order to ensure that the data on the app can be accessed across multiple devices, we need a way to store user data, perform user authentication, and ensure data security. We could use a unified backend solution such as Firebase, which includes a real time database and an authentication and security system [4]. Firestore is a noSQL database that provides faster prototyping and flexibility, allowing the data to be stored in a unified location rather than on separate devices. Firebase authentication service supports the backend of the login page and includes alternative ways to login via gmail, instagram, etc, enhancing its accessibility. Firebase also provides a security rules layer that allows us to control the degree of access that any given user is able to have on the database, ensuring data integrity. 

\begin{figure}[htbp]
    \centering
    \includegraphics[width=\columnwidth]{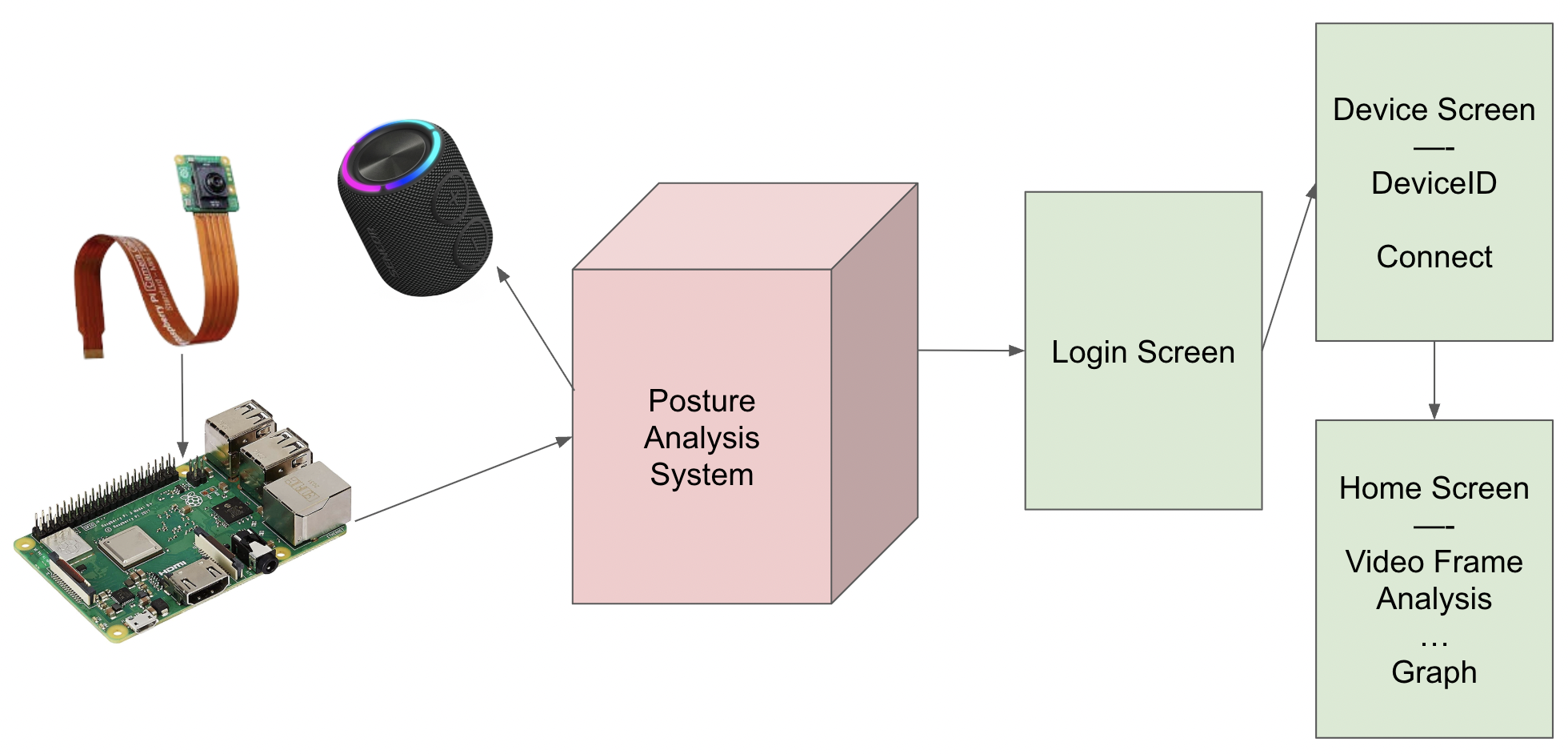}
    \caption{Flowchart of our posture monitoring system}
    \label{fig:postureapp}
\end{figure}

\begin{figure}[htbp]
    \centering
    \includegraphics[width=0.3\columnwidth]{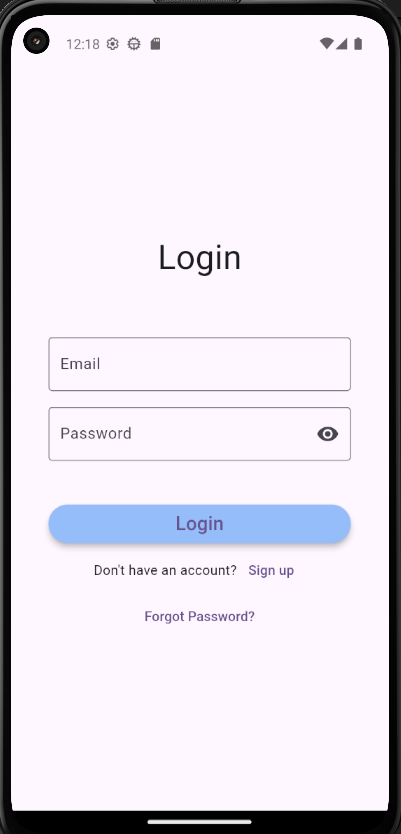}
    \caption{Login Screen of our posture monitoring system}
    \label{fig:postureapp}
\end{figure}
\begin{figure}[htbp]
    \centering
    \includegraphics[width=0.3\columnwidth]{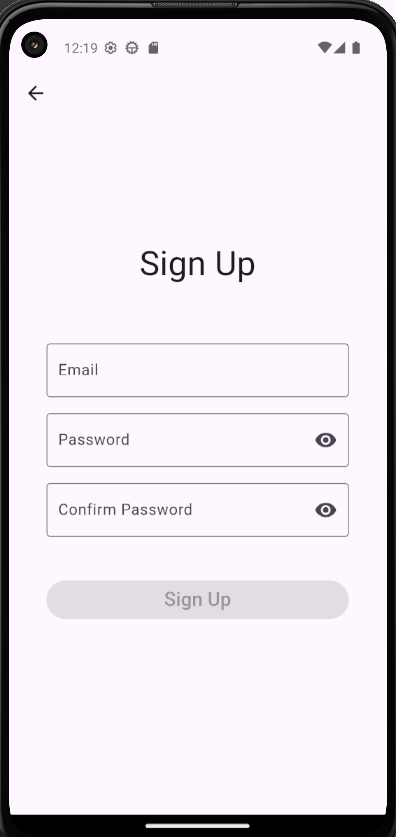}
    \caption{Signup Screen of our posture monitoring system}
    \label{fig:postureapp}
\end{figure}
\begin{figure}[htbp]
    \centering
    \includegraphics[width=0.3\columnwidth]{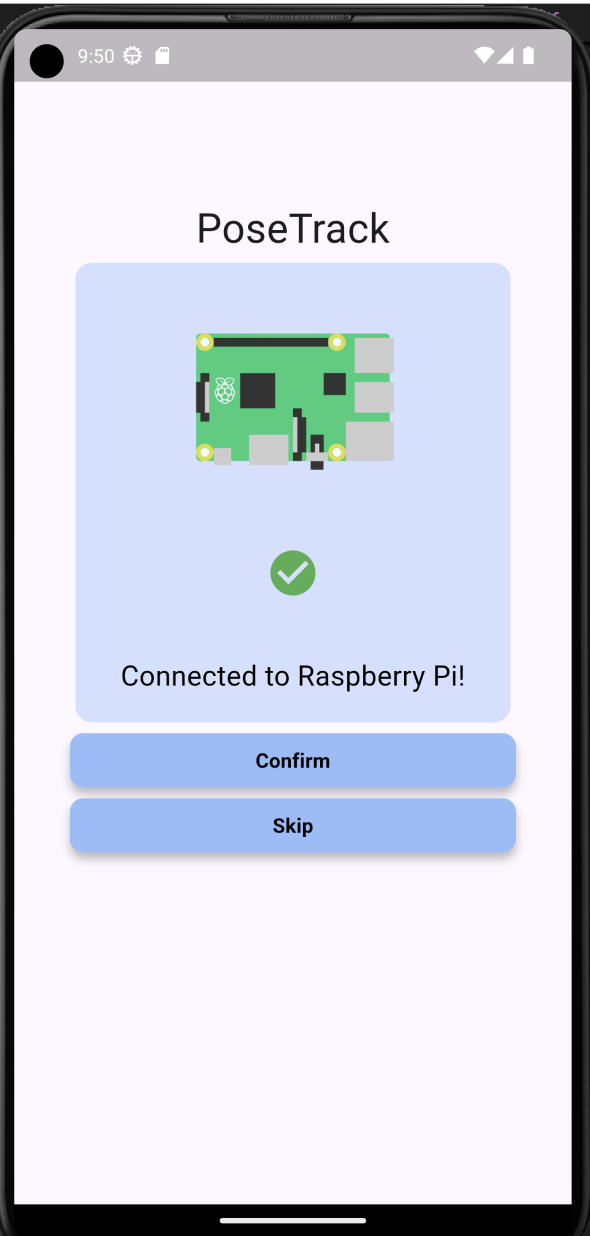}
    \caption{Device Screen of our posture monitoring system}
    \label{fig:postureapp}
\end{figure}
\begin{figure}[htbp]
    \centering
    \includegraphics[width=0.3\columnwidth]{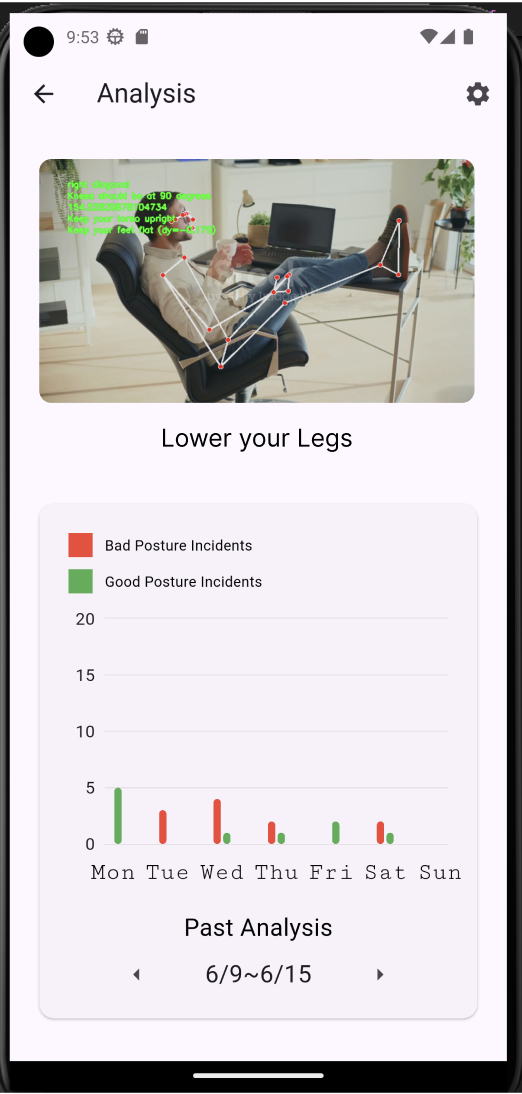}
    \caption{Home Screen of our posture monitoring system}
    \label{fig:postureapp}
\end{figure}
\vspace{7pt}
\begin{figure}[htbp]
    \centering
    \includegraphics[width=0.3\columnwidth]{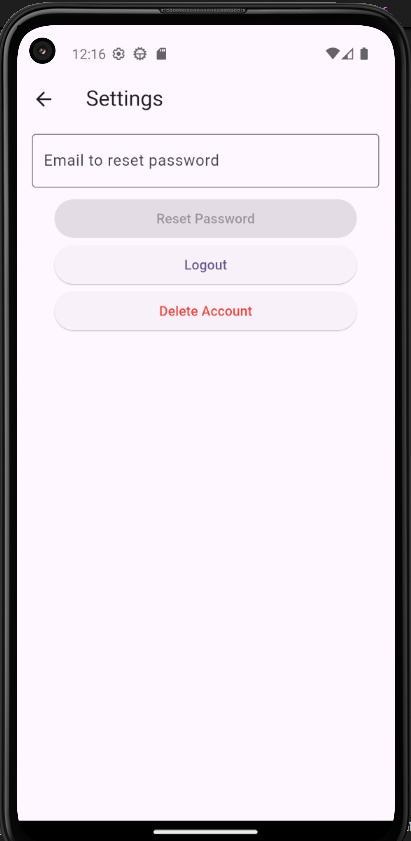}
    \caption{Settings Screen of our posture monitoring system}
    \label{fig:postureapp}
\end{figure}
\vspace{7pt}
\section{Methodology}
The main components of the program are the Raspberry Pi device, the Posture Analysis System, and the Mobile app, which work together to provide users with real-time posture feedback. A student or office worker would first set up a Raspberry Pi Camera, positioning the camera to the side or front diagonal of the desk so that the user is clearly in view. When the Raspberry Pi is connected to power, the analysis server would automatically start due to the systemd service, ensuring that the user doesn’t have to manually start the analysis engine. The posture analysis system receives the video frame from the Raspberry Pi camera, using Mediapipe Pose Landmark Detection to analyze the position of the user’s joints and determine whether the user is sitting in the correct posture. The user would then sit down at their desk and continue with their work while the Posture Analysis System runs in the background. The analysis result is constantly passed to the mobile app using a Flask server through HTTP requests. If the user crosses their legs, slouches, or performs any incorrect posture, they would see a popup appear on the app, warning them to correct their posture. The mobile app, created using the Flutter framework and programmed in the Dart language, is the frontend interface where the user can see their past data and interact with the whole system. Additionally, a speaker would play an alert sound, notifying the user to correct their posture in case they didn’t notice the popup on the app. The following section will describe each of the components in detail. \\
\subsection{Device Component}
The hardware system includes a Raspberry Pi equipped with a camera module and a speaker. The Raspberry Pi continuously captures real-time video input and analyzes the user's posture. When poor posture is detected, the speaker emits an alert sound to prompt the user to adjust their sitting position. To ensure seamless usability, we configured the system to launch automatically at startup using systemctl, eliminating the need for manual activation. The posture analysis results are transmitted to the companion mobile application via a Flask-based server, using HTTP requests for efficient communication. \\

\begin{figure}[htbp]
    \centering
    \includegraphics[width=0.8\columnwidth]{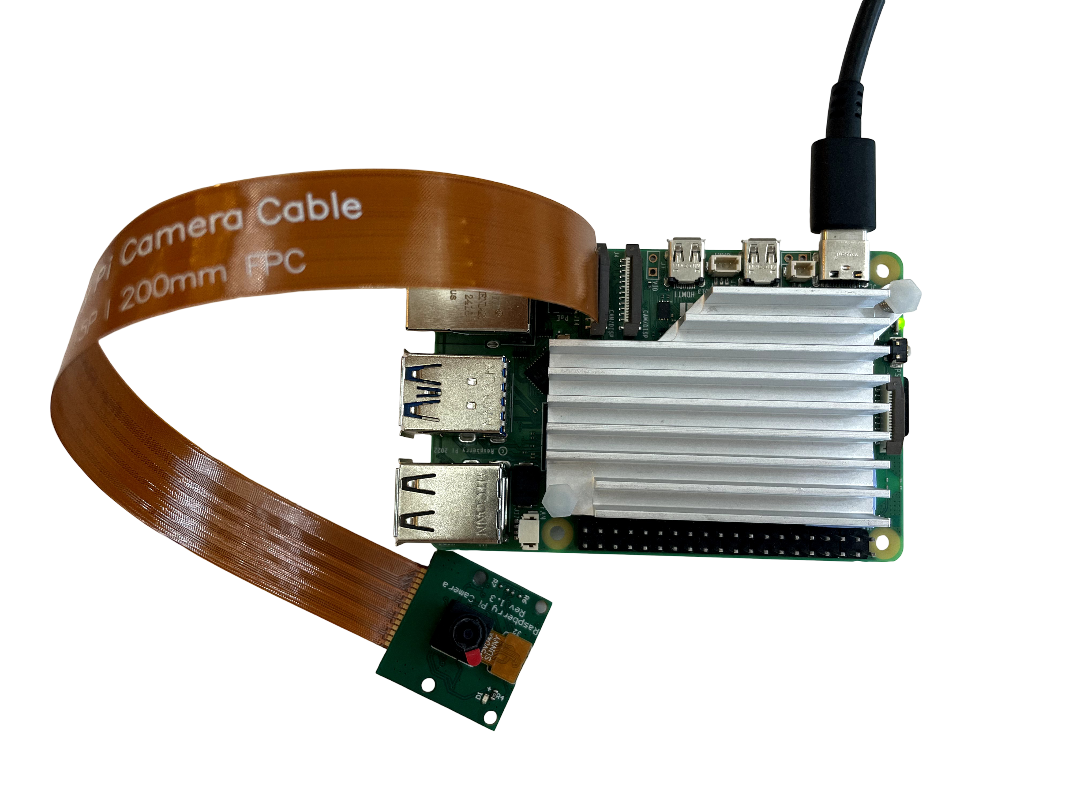}
    \caption{Prototype of the Raspberry Pi Camera System}
    \label{fig:postureapp}
\end{figure}
The function get\_posture() is used to send the Raspberry Pi video frame and the analysis result to the mobile app. get\_posture() is defined with a decorator that allows Flask to make the function respond to GET requests at the /get\_posture url endpoint. We use a thread lock when accessing the video frame to ensure that the analysis running in the background thread does not interfere with server operations. If the frame is not available, a JSON error with a 503 status code is returned. We use OpenCV to convert the frame into a JPEG buffer, then convert it to a base64 string so that data can be passed in a JSON format. If encoding to a base64 string fails, the program returns an error 500. At the end of the function, a JSON object with the frame and analysis is returned to be requested from the app.\\

We also configured a systemd service file to ensure that the Raspberry Pi server can automatically start. The [Unit] section provides a description of the service and ensures that the service only runs after network components are finished. The [Service] section defines the directory and specifies the command to start the server. Restart=on-failure ensures that systemd will automatically restart the server if it crashes or upon system boot. 

\begin{figure}[htbp]
    \centering
    \includegraphics[width=\columnwidth]{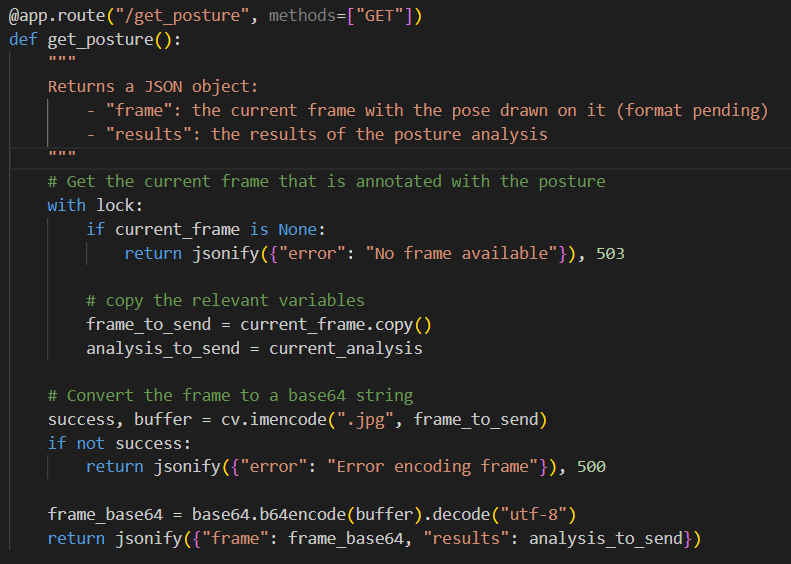}
    \caption{Code for converting the video frame into a json format to pass to the app}
    \label{fig:postureapp}
\end{figure}
\begin{figure}[htbp]
    \centering
    \includegraphics[width=\columnwidth]{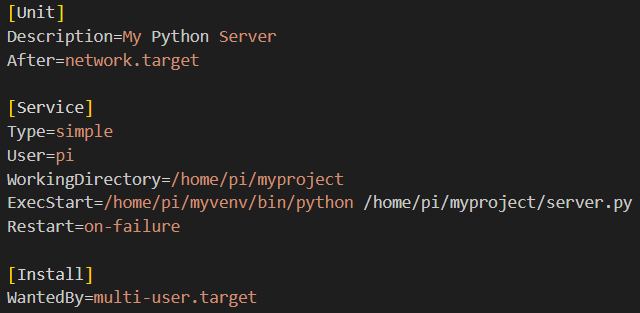}
    \caption{Code for systemctl automation}
    \label{fig:postureapp}
\end{figure}

\subsection{Analysis System}
The Posture Analysis System analyzes video data from the Raspberry Pi and corrects the user’s posture. The program uses Mediapipe posture landmark detection to locate the user’s body landmarks from the video frame. Then the landmarks are used to check for bad posture, such as if the user crossed their legs, leaned back or forward, or raised their feet above their hips, etc.\\

\begin{figure}[htbp]
    \centering
    \includegraphics[width=0.7\columnwidth]{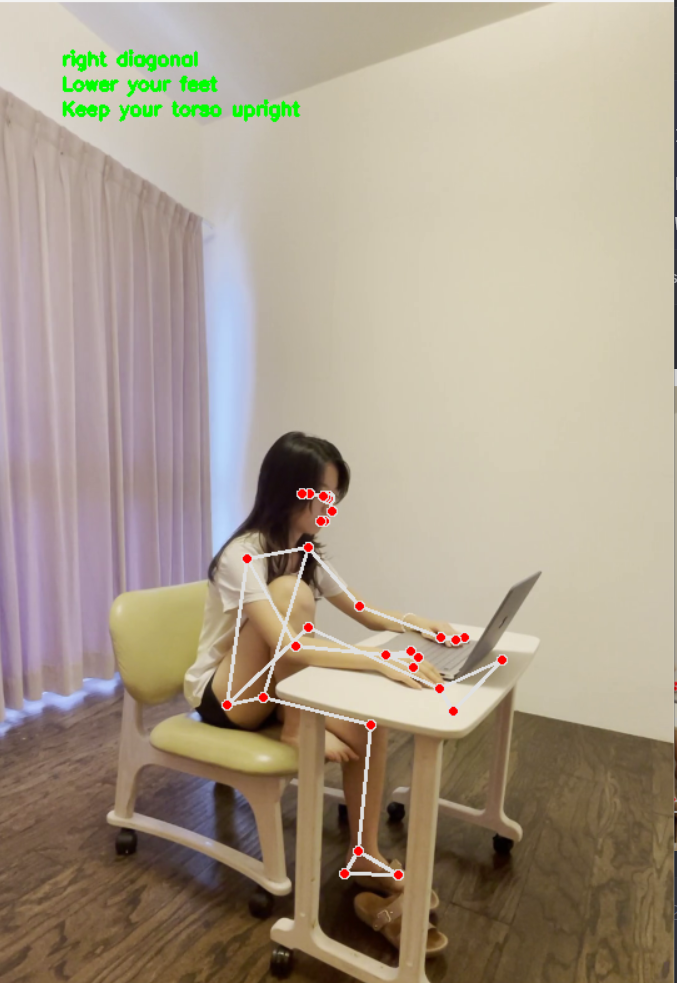}
    \caption{Sample image of Analysis Frame and Results}
    \label{fig:postureapp}
\end{figure}
\begin{figure}[htbp]
    \centering
    \includegraphics[width=\columnwidth]{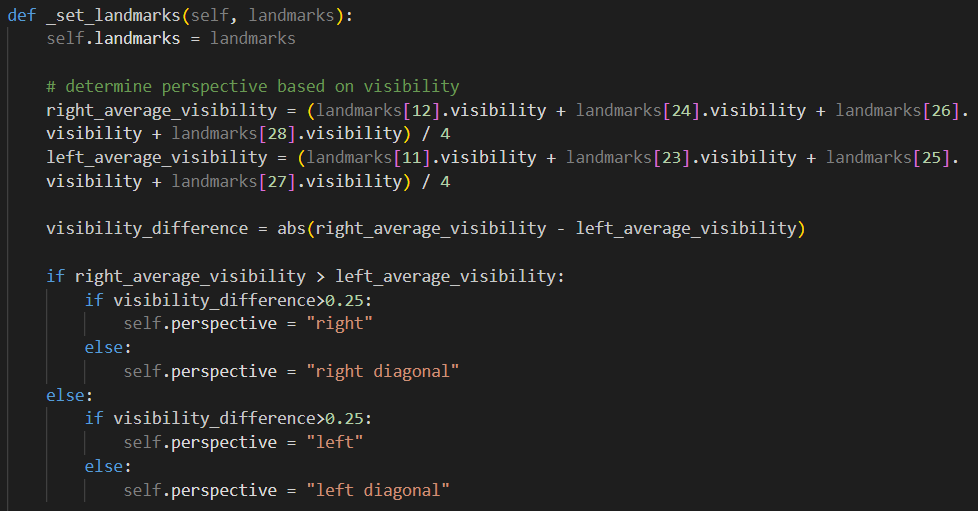}
    \caption{Code to determine the user's Perspective}
    \label{fig:postureapp}
\end{figure}
\begin{figure}[htbp]
    \centering
    \includegraphics[width=\columnwidth]{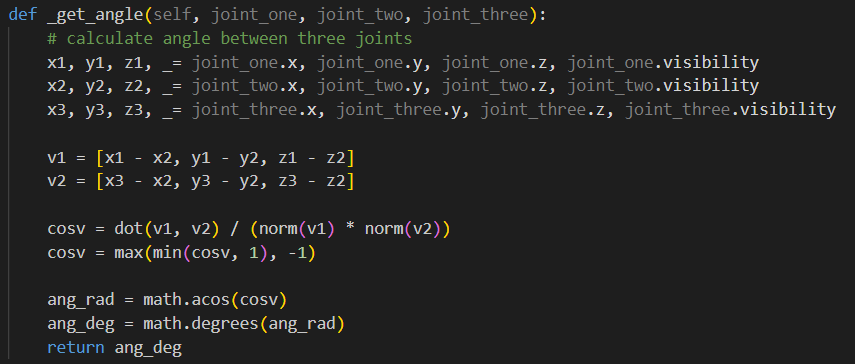}
    \caption{Code to get the Angle between three Joints}
    \label{fig:postureapp}
\end{figure}
The function set\_landmarks() first calculates the perspective that the camera is positioned relative to the user based on the left and right average visibility. From the visibility difference between the left and right sides, the perspective is categorized as either left, right, left diagonal, or right diagonal. The perspective will determine which landmarks we use to analyze the user’s posture. \\

The function get\_angle() calculates the angle between three joints. The coordinates are used to calculate the two vectors that form the angle, then the angle is calculated based on the dot product and the magnitude of the two vectors. get\_angle() will be repeatedly used in the analysis code to check for criteria such as whether the user is sitting straight, if knees are around 90 degrees, or if the user is leaning their head too far forward or back, etc.\\  

The functions check\_leg\_crossed(), check\_back\_angle(), check\_feet\_above\_hips(), etc, all check for one criterion of good posture. These functions use the perspective determined in set\_landmarks() to decide which landmarks to use, then use get\_angle() to check if a certain posture is inaccurate. For instance, the function checking if the user is leaning back uses the angle between knees, hips, and shoulders to determine if the user is leaning front or back. The other functions all perform similar tasks of checking one of the criteria.

\subsection{Mobile App}
The mobile app consists of a login/signup screen, device screen, home screen, and a settings screen. The user is first brought to the login/signup screen. After they logged into their account, the user would be brought to the device connection page, where they would have a choice to connect to the Raspberry Pi or skip the connection and only view past data. The user would then be shown the home screen, which displays the current video input, analysis, and a graph with the past posture data. In the home screen, the users have the option to go to the settings page which allows them to log out, change their password, or delete their account. 
\begin{figure}[htbp]
    \centering
    \includegraphics[width=\columnwidth]{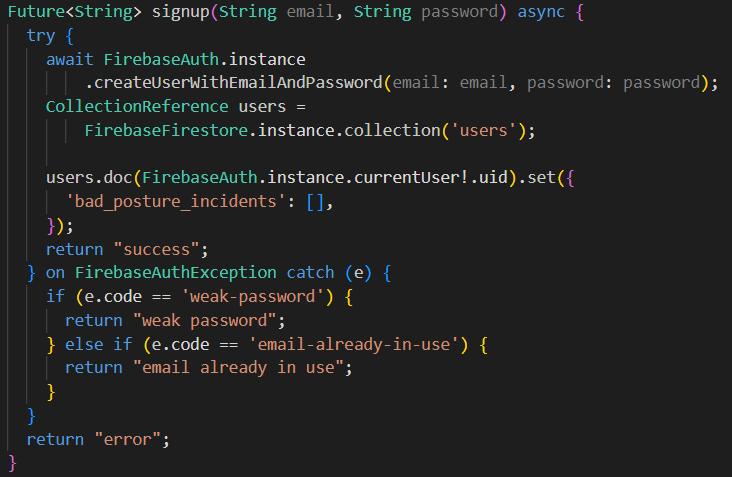}
    \caption{Code implementing the backend data storage connecting the app to Firebase}
    \label{fig:postureapp}
\end{figure}

The login/signup screen serves as the front end of the authentication system while Firebase Authentication serves as the backend. The code defines an asynchronous Dart function called \texttt{signup()} that registers new users and stores user data in Firestore. New users are initialized with the Firebase \texttt{createUserWithEmailAndPassword method}. When the user is registered, an empty \texttt{bad\_posture\_incidents} list is stored in Firestore under the user’s ID. The function also checks for several problems such as a weak password or email address already in use. This code block connects the frontend user interface with data storage, allowing the data to be properly passed through the app to Firebase. 

\begin{figure}[htbp]
    \centering
    \includegraphics[width=\columnwidth]{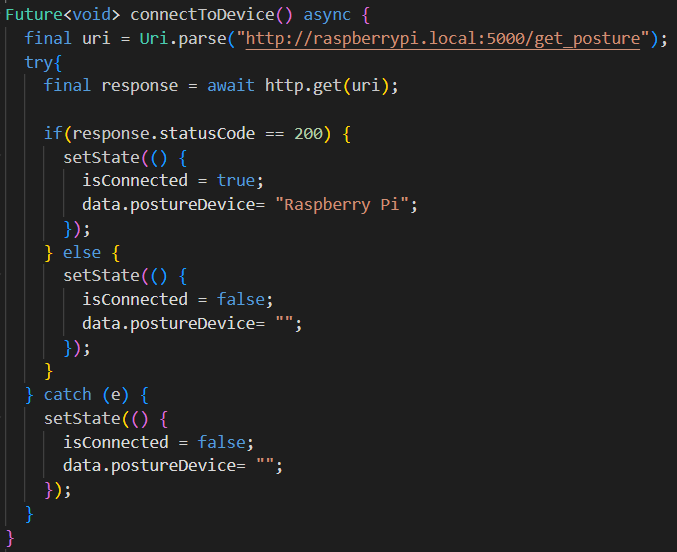}
    \caption{Code implementing device connection to the Raspberry Pi}
    \label{fig:postureapp}
\end{figure}
The function \texttt{connectToDevice()} checks if the Raspberry Pi device is connected to the app. The function sends out a GET request to the Raspberry Pi server at the “http://raspberrypi.local:5000/get\_posture” endpoint. If the status code is 200, the connection was successful, and the user would be allowed to press the confirm button. We opted to use a local server endpoint as it is both cheaper and more private. A local server means that there is no need to run a server on our end which saves on cost and latency while also never requiring users to send any recording of themselves anywhere other than the device.\\

To manage user data throughout the app, we used a singleton Data class inheriting from Flutter’s ChangeNotifier. This class is responsible for storing and broadcasting changes to a central instance while communicating with the top-level StreamBuilder widget, which is in charge of getting data from the Firestore database. It also includes the ability to write to Firestore. Once any data is modified, the method \texttt{notifyListeners()}, provided by {ChangeNotifier, is used to prompt relevant widgets to update themselves. The Data class connects the data inputted into the UI to Firestore, and is key to ensure the App reflects real time data. 
\vspace{7pt}
\section{Experiment}
\subsection{Experiment 1: Analysis Accuracy}
One essential function of our system is to accurately detect whether the user has poor posture. Validating the accuracy of the analysis system is critical because wrong analysis results would result in false warnings. To test the posture detection accuracy, videos of subjects were recorded sitting in four postures that include correct posture, crossed legs, leaned forward, and legs on a chair. Videos of the four postures are recorded from all four perspectives, left, right, left diagonal, and right diagonal. Each action in the video is maintained for at least 10 seconds to ensure that the system can clearly detect it. Several environmental factors that could affect the experiment were also kept as controls. The videos were recorded in the same lighting, background, and with the same clothing throughout all videos. Then, we manually labeled each video with their corresponding posture and compared it to the analysis system results.\\ 
\begin{figure}[htbp]
    \centering
    \includegraphics[width=0.7\columnwidth]{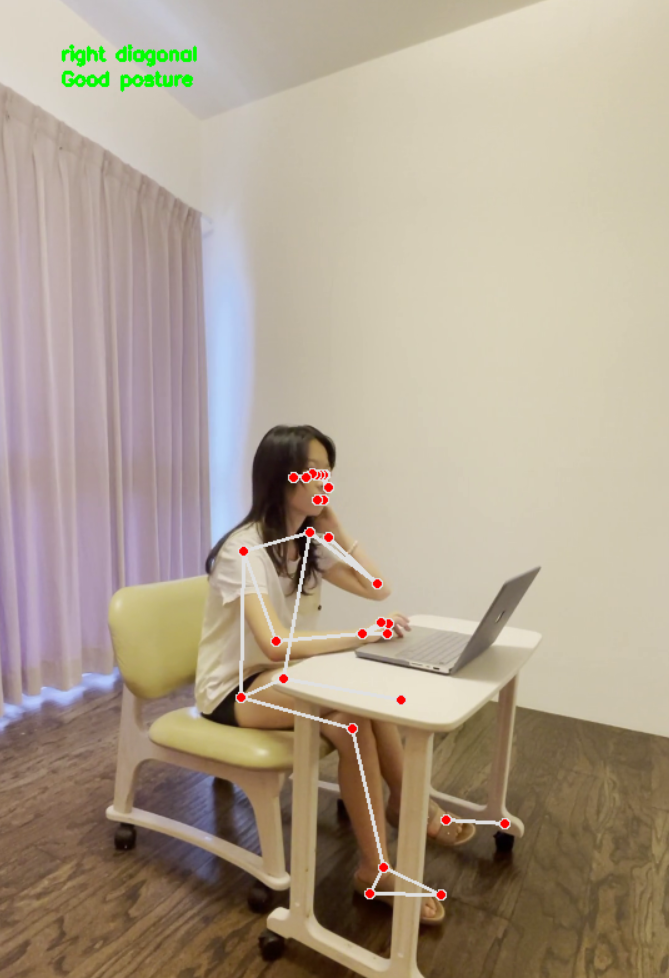}
    \caption{Incorrect analysis of crossed legs due to legs being blocked by table}
    \label{fig:postureapp}
\end{figure}

\begin{figure}[htbp]
    \centering
    \includegraphics[width=0.7\columnwidth]{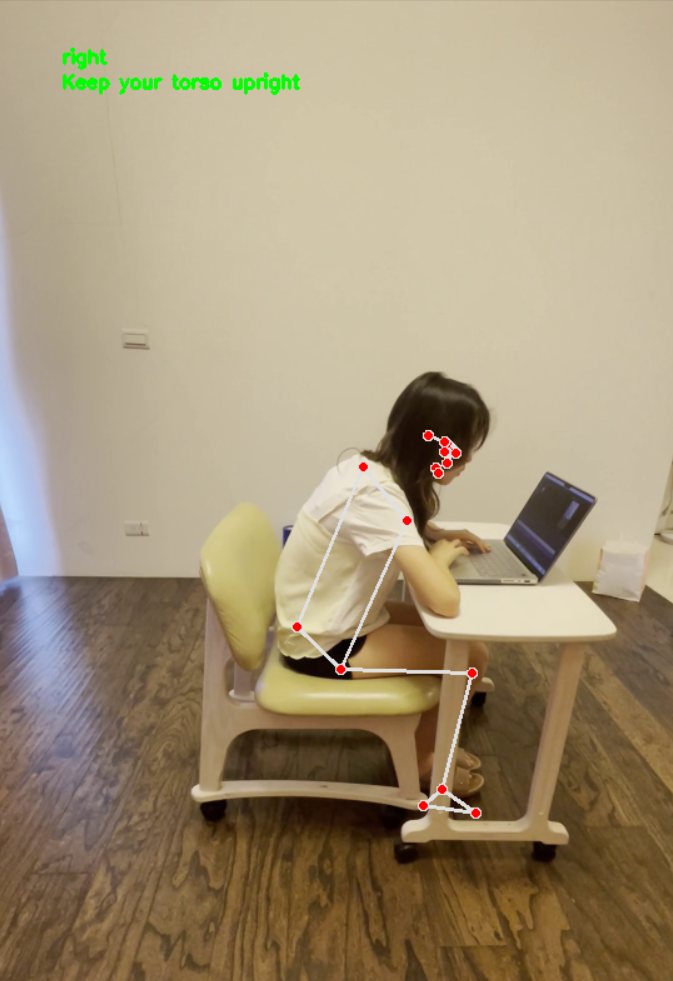}
    \caption{Correct Detection of Forward Lean}
    \label{fig:postureapp}
\end{figure}

From the videos tested, the accuracy for identifying good posture and leaning forward is 70\% and 100\%, respectively. In contrast, the system failed to identify crossed legs in all the videos and only identified the leg-on-chair posture for  50\% of the videos. The low accuracy for crossed legs and leg-on-chair tests is due to the knees often being blocked by the table in the diagonal perspective and the arms in the left/right perspective. As a result, Mediapipe is often unable to correctly label the positions of the user’s knees, leading to inaccurate results. Nevertheless, for the postures that Mediapipe is able to correctly label, such as leaning forward and good posture, the analysis system produced results with high accuracy. Despite the low accuracy for identifying leg positions, the results show that the analysis system has high accuracy for postures that are clearly visible. The camera angle and view can be further adjusted to ensure that the user’s leg positions are fully visible to the analysis system. 

\begin{table}[htbp]
\centering
\resizebox{\linewidth}{!}{%
\begin{tabular}{|lcccc|}
\hline
\textbf{Perspective} & \textbf{Good} & \textbf{Lean} & \textbf{Crossed} & \textbf{Legs on} \\
                     & \textbf{Posture} & \textbf{Forward} & \textbf{Legs} & \textbf{Chair} \\
\hline
Right Diagonal & \ding{51} & \ding{51} & \ding{55} & \ding{51} \\
Right          & \ding{51} & \ding{51} & \ding{55} & \ding{55} \\
Left Diagonal  & \ding{55} & \ding{51} & \ding{55} & \ding{51} \\
Left           & \ding{51} & \ding{51} & \ding{55} & \ding{55} \\
\hline
\end{tabular}%
}
\caption{Posture Detection Accuracy Across Perspectives}
\label{tab:posture_accuracy}
\end{table}

\subsection{Experiment 2: Lighting Conditions}
Environmental factors such as lighting may also impact the accuracy of the analysis system. Therefore, we designed an experiment to determine the effects of these factors in varying degrees. We tested the system on videos of a user sitting in correct posture and leaning forward. Videos of the two postures are recorded from all four perspectives: left, right, left diagonal, and right diagonal. The lighting for the videos was then adjusted to five levels of brightness: very dim, dim, normal, bright, and very bright. The adjusted videos were then given to the analysis system, and the result of the analysis was compared with manual labels to compare how lighting affected the analysis. The user wore the same clothes and sat in the same environment to ensure that other factors were all kept in control. We also used the same pose sequences and only adjusted the lighting, which ensures that only the lighting contributes to the result. 
\begin{figure}[htbp]
    \centering
    \includegraphics[width=\columnwidth]{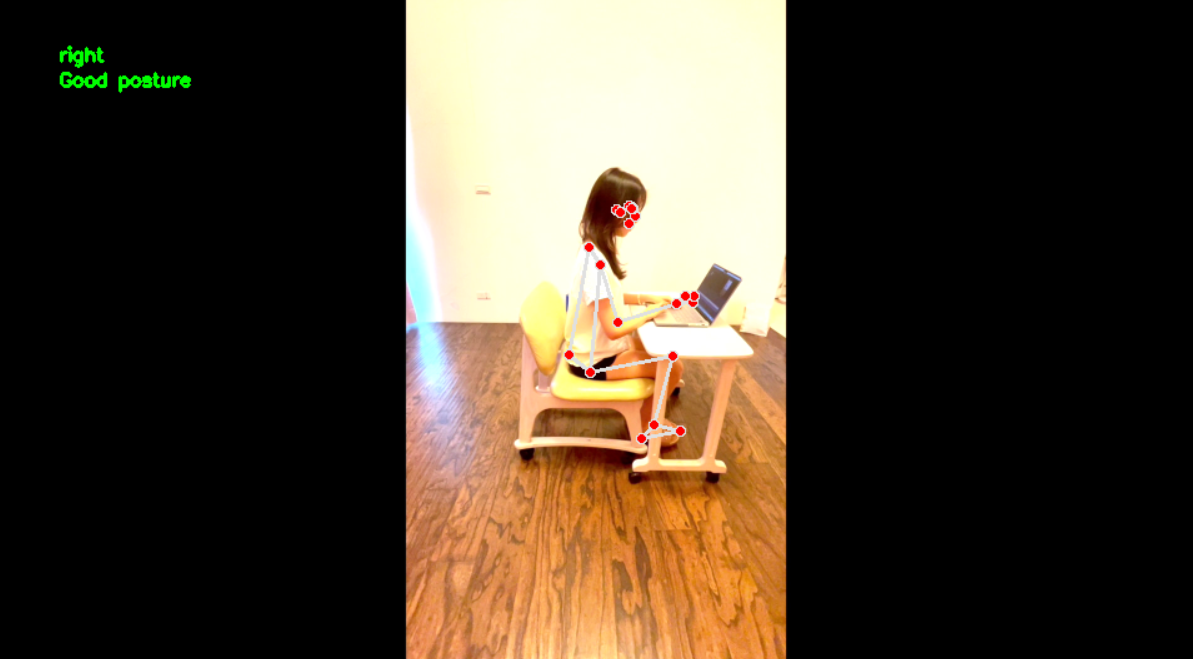}
    \caption{Correct Detection of Good Posture in Bright Lighting}
    \label{fig:postureapp}
\end{figure}
\begin{figure}[htbp]
    \centering
    \includegraphics[width=\columnwidth]{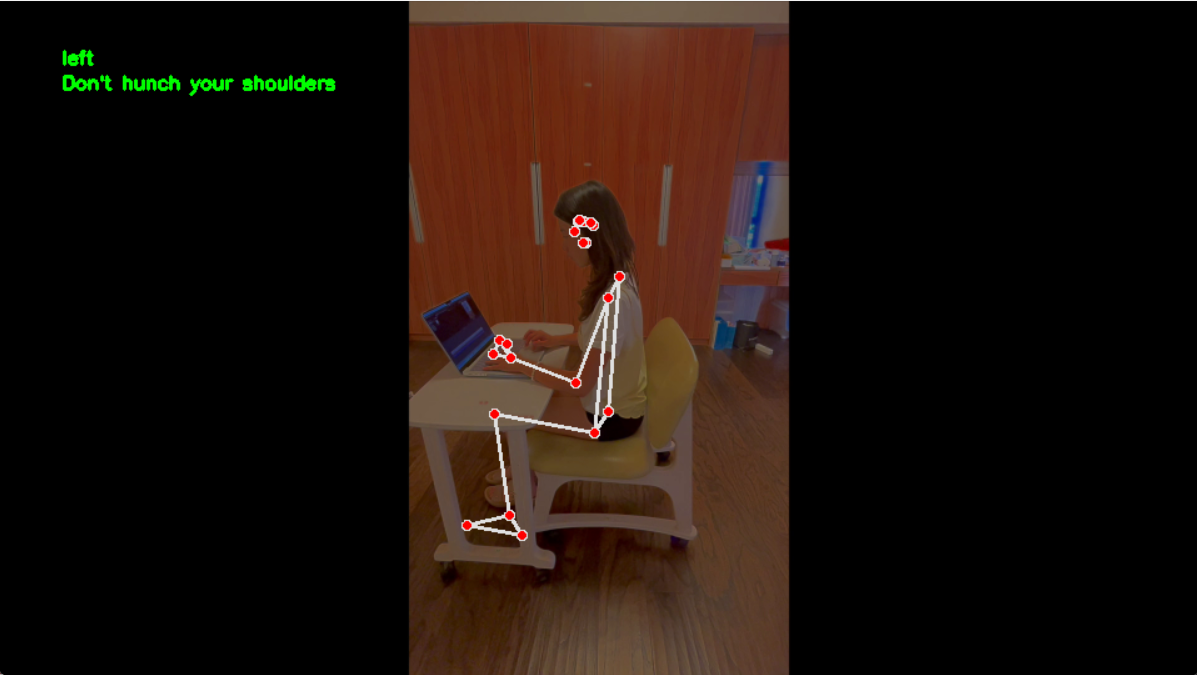}
    \caption{Incorrect Detection of Good Posture in Dim Lighting}
    \label{fig:postureapp}
\end{figure}
\begin{table}[htbp]
\centering
\begin{tabular}{|lccccc|}
\hline
{} &    Very Dim &         Dim &      Normal &      Bright & Very Bright \\
\hline
Right Diagonal &  \ding{51} &  \ding{51} &  \ding{51} &  \ding{51} &  \ding{51} \\
Right          &  \ding{51} &  \ding{51} &  \ding{51} &  \ding{51} &  \ding{51} \\
Left Diagonal  &   \ding{55} &   \ding{55} &   \ding{55} &   \ding{55} &   \ding{55} \\
Left           &   \ding{55} &  \ding{51} &  \ding{51} &  \ding{51} &  \ding{51} \\
\hline
\end{tabular}

\caption{Posture Detection Accuracy for Good Posture under Varying Lighting Conditions}
\label{tab:good_posture_lighting}
\end{table}

\begin{table}[htbp]
\centering
\begin{tabular}{|lccccc|}
\hline
{} &    Very Dim &         Dim &      Normal &      Bright & Very Bright \\
\hline
Right Diagonal &  \ding{51} &  \ding{51} &  \ding{51} &  \ding{51} &  \ding{51} \\
Right          &  \ding{51} &  \ding{51} &  \ding{51} &  \ding{51} &   \ding{55} \\
Left Diagonal  &   \ding{55} &   \ding{55} &  \ding{51} &  \ding{51} &  \ding{51} \\
Left           &  \ding{51} &  \ding{51} &  \ding{51} &  \ding{51} &  \ding{51} \\
\hline
\end{tabular}

\caption{Posture Detection Accuracy for Lean Forward under Varying Lighting Conditions}
\label{tab:lean_forward_lighting}
\end{table}

The results of the experiment revealed high accuracies across most lighting conditions. It achieved high accuracies of 100\% for good posture detection in right and right diagonal perspectives, a moderately high accuracy of 80\% for left perspective, but a low accuracy of 0\% for the left diagonal perspective. The low accuracy for the left diagonal video is due to the legs being obscured by the desk, leading to a blind spot that exists despite the lighting conditions. The results for forward lean also have high accuracy. The system accurately detected forward lean in all the videos except one very dim video for the left diagonal perspective, and one very bright video for the right perspective, achieving an accuracy of 90\%. Overall, the results demonstrate that the system is resilient to environmental change and is more affected by the camera view, as it can still accurately detect good posture and forward lean even with changes in lighting.

\vspace{7pt}
\section{Related work}
Tlili et al. (2021) \cite{Tlili} created a system to address neck and back pains caused by slouching. The researchers designed a belt equipped with inertial sensors to strap to the user’s back and collect data. The data is sent to a cloud server to be processed and the results are sent to the mobile application. The system measures the user’s slouching angle and sends an alert if the user's slouching degree exceeds 30 degrees for over 5 minutes. While the solution is effective in determining whether the user is slouching, it doesn’t consider other factors such as crossed legs, forward or backward leaning, or head position. Our project improves on this by considering all these factors and eliminating the need for a wearable device.\\

Gupta et al. (2021) \cite{Gupta} developed a posture detection system that uses 3 accelerometers placed on the user’s back to detect poor posture. The researchers used a one class support vector machine and isolation forest along with an unsupervised k-means clustering algorithm to detect poor posture. The results showed that the isolation forest provides the highest accuracy of 99.3\%. Despite the high accuracy, the method involves placing 3 accelerometers on the user’s back, which makes it less likely to be widely used outside the laboratory environment. Our solution improves on this method by eliminating the need for a wearable device, making posture correction more widely accessible. 

Jang et al. (2020) \cite{Jang} proposed a system using accelerometers and magnetometers to detect poor postures, including forward head, rounded shoulders, and elevated shoulders. The researchers processed the sensor data using a deep neural network (DNN) and a convolutional neural network (CNN). The results demonstrated successful bad posture classification with high accuracy of 88.1\% and 88.7\% for DNN and CNN respectively. Our solution does not use any bespoke neural networks but uses Mediapipe pose landmarks to analyze the user’s posture. Our solution encompasses checks for a wider range of poor postures to include analyzing leg positions instead of only head and shoulder positions as in this study.
\vspace{7pt}
\section{Conclusion and future work}
Currently, the app provides users with general posture warnings and tracks weekly posture data over time. To enhance its effectiveness, future versions could offer personalized recommendations tailored to users' individual posture habits. The Posture Analysis System can also be expanded to include additional evaluation criteria, such as head tilt (left/right), shoulder alignment, and distance from the computer, which would further improve detection accuracy. One current limitation lies in the hardware setup—the Raspberry Pi camera is still in a prototype stage and lacks a rigid mounting structure, which may result in inconsistent data capture. This can be addressed by designing a stable enclosure or adjustable mount to ensure reliable positioning and data quality. Furthermore, the app can be extended beyond posture correction by incorporating wellness features, such as reminders to take breaks, stretch, or perform light exercises. These enhancements would not only make the system more robust but also support a more holistic approach to long-term ergonomic health.\\

PoseTrack demonstrates the ability to use computer vision and affordable hardware to promote a healthier lifestyle. Through combining Mediapipe Pose Detection with an interactive Flutter based mobile application, the system creates an accessible solution for monitoring poor posture. While there are limitations in accurately detecting leg postures under cases of low visibility, the system performed well for other body joints that are fully visible. In the future, we can continue to make improvements and refine the posture detection system. 
\vspace{7pt}



%

\end{document}